\documentclass{ws-procs9x6}

\begin{document}

\title{
ELECTROPRODUCTION OF RESONANCES \\
AT LARGE MOMENTUM TRANSFERS}

\author{V. M. BRAUN}

\address{Institut f\"ur Theoretische Physik, Universit\"at Regensburg, \\
         93040 Regensburg, Germany\\[1mm]
E-mail: vladimir.braun@physik.uni-regensburg.de}

\begin{abstract}
Measurements of transition form factors for the electro-excitation 
of nucleon resonances in the $Q^2=5-14$~GeV$^2$ region can provide one
with the information on quark wave functions at small transverse separations.
In particular a comparison of form factors for the states of opposite parity
can give insight in the mechanisms of chiral symmetry breaking.   
I discuss perspectives of the theoretical description of such reactions 
using a combination of lattice calculations and light-cone sum rules.
\end{abstract}


\bodymatter

\section{Introduction}\label{aba:sec1}

Electroproduction of nucleon resonances has long been recognized as 
an important tool in the exploration of the nucleon structure at 
different scales. 
There is a growing consensus that perturbative QCD (pQCD) factorization 
based on hard gluon exchange is not reached at present energies; 
however, the emergence of quarks and gluons as the adequate degrees of 
freedom is expected to happen earlier, 
at $Q^2\sim$ a few GeV$^2$. Measurements of the
form factors in this transition region are planned at Jefferson Lab 
\cite{Burkert:2008mw,Aznauryan:2009da} using the CLAS12 detector. In this talk 
such reactions are addressed from the theory perspective.
I try to formulate the physics goals and explain an approach to
the electroproduction of resonances that combines lattice calculations
of wave functions at small transverse separations with dispersion relations 
and quark-hadron duality, known as light-cone sum rules (LCSR).                

Quantum chromodynamics (QCD) predicts  
\cite{Chernyak:1977as,Efremov:1979qk,Lepage:1979za,Lepage:1980fj}
that at large momentum transfer the transition form factors become increasingly 
dominated by the contribution of the valence Fock state with small transverse 
separation between the partons. In reality the dominance of valence configurations 
is likely to be achieved but ``shrinkage'' to small transverse
separations does not seem to occur and also power counting rules based on helicity 
conservation do not work at moderate $Q^2$ that we are able to study experimentally.

To be more specific, the nucleon valence state contains contributions  with 
different orbital angular momentum $\ell_z=0,\pm 1$. The leading contribution to
form factors at large $Q^2$ comes from the $\ell_z=0$ state\cite{Lepage:1980fj}:
\begin{eqnarray*}
 |P\uparrow\rangle^{\ell_z=0} &=& \int\frac{[dx][d^2\vec{k}]}{6\sqrt{x_1x_2x_3}}
\,\,{\psi^{\ell_z=0}(x_i,\vec{k}_i)}\times\\
&&\hspace*{-1.2cm}\times
\Big\{
 \big|u^\uparrow(x_1,\vec{k}_1)u^\downarrow(x_2,\vec{k}_2)d^\uparrow(x_3,\vec{k}_3)\big\rangle
-\big|u^\uparrow(x_1,\vec{k}_1)d^\downarrow(x_2,\vec{k}_2)u^\uparrow(x_3,\vec{k}_3)\big\rangle
 \Big\}
\end{eqnarray*} 
where $\psi^{\ell_z=0}(x_i,\vec{k}_i)$ is the three-quark light-cone wave function 
that depends on quark momentum fractions $x_i$ and transverse momenta $\vec{k}_i$%
\footnote{Here and below we do not show contributions which vanish in the limit 
 of small transverse separations, cf. Ref.~[\refcite{Ji:2002xn}].}.

The simplification that occurs at asymptotically large $Q^2$ is that the 
$\vec{k}$--dependence of wave functions becomes irrelevant and all necessary 
(nonperturbative) information is contained in the integral over transverse 
momenta
\begin{eqnarray*}
{\Phi_3(x_i;\mu)} &=& 
 \int^\mu [d^2\vec{k}]\,\, \psi^{\ell_z=0}(x_i,\vec{k}_i) 
 \end{eqnarray*}
where the cutoff $\mu\sim Q$ has to be imposed to make the integral converge. 

The function $\Phi_3(x_i;\mu)$ is called the leading-twist distribution amplitude (DA).
It can be studied using the operator product expansion\cite{Braun:2008ia}
\begin{eqnarray}
\Phi_3 (x_i;\mu) &\! =\! & 120 {f_N}(\mu) x_1 x_2 x_3 
 \Big\{  1     + {c_{10}}(\mu) (x_1 \! - 2 x_2 +x_3)
+ {c_{11}}(\mu) (x_1  \!-\! x_3)
\nonumber\\ &&{}  
 + {c_{20}}(\mu) \left[ 1 + 7 (x_2 - 2 x_1 x_3 - 2 x_2^2) \right]
+ {c_{21}}(\mu)\left( 1 - 4 x_2 \right) \left( x_1 \!-\! x_3 \right)
\nonumber\\&&{} 
 \left. + {c_{22}}(\mu)
\left[ 3 - 9 x_2 + 8 x_2^2 - 12 x_1 x_3 \right]+\ldots
\right\}
\label{twist3}
\end{eqnarray}
where $f_N(\mu)$ (wave function at the origin) and $c_{ik}(\mu)$ (shape parameters) 
are scale-dependent coefficients which
can be defined as matrix elements of (multiplicatively renormalizable)
local operators. They can be calculated using QCD sum rules\cite{Chernyak:1983ej} 
or lattice QCD\cite{Braun:2008ur}. The DA $\Phi_3(x_i;\mu)$ is thus a much simpler
object compared to the full light-cone wave function $\psi^{\ell_z=0}(x_i,\vec{k}_i)$;
unfortunately this reduction does not seem to work in the $Q^2=5-15$~GeV$^2$ range.

Another problem is that the standard pQCD factorization approach does not take
into account contributions of states with non-vanishing orbital angular momentum.
{}For example\cite{Ji:2002xn}
\begin{eqnarray*} 
 |P\uparrow\rangle^{\ell_z=1} &=& \int\frac{[dx][d^2\vec{k}]}{6\sqrt{x_1x_2x_3}}
\,\,\Big[
k_1^+{\psi_1^{\ell_z=1}(x_i,\vec{k}_i)} + k_2^+{\psi_2^{\ell_z=1}(x_i,\vec{k}_i)}
    \Big]\times\\
&&\hspace*{-1.2cm}\times
\Big\{
 \big|u^\uparrow(x_1,\vec{k}_1)u^\downarrow(x_2,\vec{k}_2)d^\downarrow(x_3,\vec{k}_3)\big\rangle
-\big|d^\uparrow(x_1,\vec{k}_1)u^\downarrow(x_2,\vec{k}_2)u^\downarrow(x_3,\vec{k}_3)\big\rangle
 \Big\}
\end{eqnarray*}
where $k^\pm = k_x\pm i k_y$.
The new light-cone wave functions ${\psi_1^{\ell_z=1}(x_i,\vec{k}_i)}$
and ${\psi_2^{\ell_z=1}(x_i,\vec{k}_i)}$ are reduced to next-to-leading twist-4 nucleon 
DAs~\cite{Braun:2000kw,Belitsky:2002kj}
\begin{eqnarray*}
{\Phi_4(x_i;\mu)} &=& \int^\mu \frac{[d^2\vec{k}]}{m_N x_3}\,\,
 k_3^-\big[k_1^+{\psi_1^{L=1}} + k_2^+{\psi_2^{L=1}}\big](x_i,\vec{k}_i)
\nonumber\\
{\Psi_4(x_i;\mu)} &=& \int^\mu \frac{[d^2\vec{k}]}{m_N x_2}\,\,
 k_2^-\big[k_1^+{\psi_1^{L=1}} + k_2^+{\psi_2^{L=1}}\big](x_i,\vec{k}_i)
\end{eqnarray*} 
and, again, can be studied using OPE\cite{Braun:2000kw,Braun:2008ia}
\begin{eqnarray}
 \Phi_4(x_i;\mu) &=& 12{\lambda_1}(\mu)x_1 x_2 
                +  12{f_N}(\mu)x_1 x_2\left[1+\frac23(1-5x_3)\right]+\ldots 
\nonumber\\
 \Psi_4(x_i;\mu) &=& 12{\lambda_1}(\mu)x_1 x_3 
                +  12{f_N}(\mu)x_1 x_3\left[1+\frac23(1-5x_2)\right]+\ldots 
\label{twist4}
\end{eqnarray}
Note that to this accuracy twist-4 DAs include one new parameter only, $\lambda_1$.
A similar expansion and the reduction to DAs can be worked out for nucleon resonances.
Electroproduction of nucleon resonances at high $Q^2$ gives access to three-quark wave 
functions, more precisely to the overlap integrals between the wave functions of the 
nucleon and the resonance. These overlap integrals are in general too complicated 
to be calculated in QCD directly and our strategy will be to reduce these integrals to
convolutions of distribution amplitudes which can be constrained using lattice QCD.   

We believe that one important physics goal for such studies will be to compare 
DAs (alias light-cone wave functions at small transverse separations) of baryon
states with opposite parity, e.g. $J^P=1/2^+$ and $J^P=1/2^-$. It is well known that 
such ``parity-doublets'' are non-degenerate in QCD because of spontaneous 
breaking of chiral symmetry. It is not known, however, whether this difference mainly
affects the ``pion cloud'' or it is present at short distances already and affects 
momentum fraction distributions of valence quarks. Moreover, there are 
indications\cite{Glozman:2007ek} that chiral symmetry is effectively restored 
in the spectra of higher-mass resonances and it would be extremely interesting to compare
the corresponding wave functions. 

The general strategy of combining the constraints on DAs from a lattice calculation 
with LCSRs to calculate the form factors is suggested in Ref.~[\refcite{Braun:2009jy}]
where, as the first demonstration, we considered the electroproduction of 
$N^*(1535)$, the parity partner of the nucleon.
In what follows I describe an ongoing work in this 
direction by the Regensburg theory group and QCDSF collaboration.

\section{Nucleon and $N^*(1535)$ Distribution Amplitudes from Lattice QCD}

Baryon states of different parity can be identified in a lattice calculation as 
those propagating forward or backward in (imaginary) time \cite{Lee:1998cx,Sasaki:2001nf},
so in fact the results for $N^*(1535)$ reported in Ref.~[\refcite{Braun:2009jy}]
are essentially a byproduct of our calculation of the nucleon 
DAs\cite{Braun:2008ur}. These results were obtained using QCDSF/DIK gauge configurations 
with two flavors of clover fermions for two different $\beta$ values and several
quark masses on $24^3\times 48$ lattices. The calculation was done using 
nonperturbatively renormalized three-quark operators\cite{Gockeler:2008we}
 with up to two derivatives
imposing a RI'-MOM--like renormalization condition and converting the results to the 
$\overline{MS}$ scheme. In this way the mixing with ``total derivatives'' is automatically 
taken into account. 

This work will be  continued, using larger lattices and smaller pion masses in order
to minimize effects of the chiral extrapolation. In Fig.~1 I present preliminary 
results\cite{schiel} of the new calculation using ca. 600 $N_f=2$ gauge 
configurations on a $32^3\times 64$ lattice with $\beta=5.29$ (a = 0.0753 fm) 
and pion mass $m_\pi \simeq 282$ MeV ($m_\pi L = 3.44)$).  
\begin{figure}[t]
  \includegraphics[width=0.385\textwidth,angle=-90]{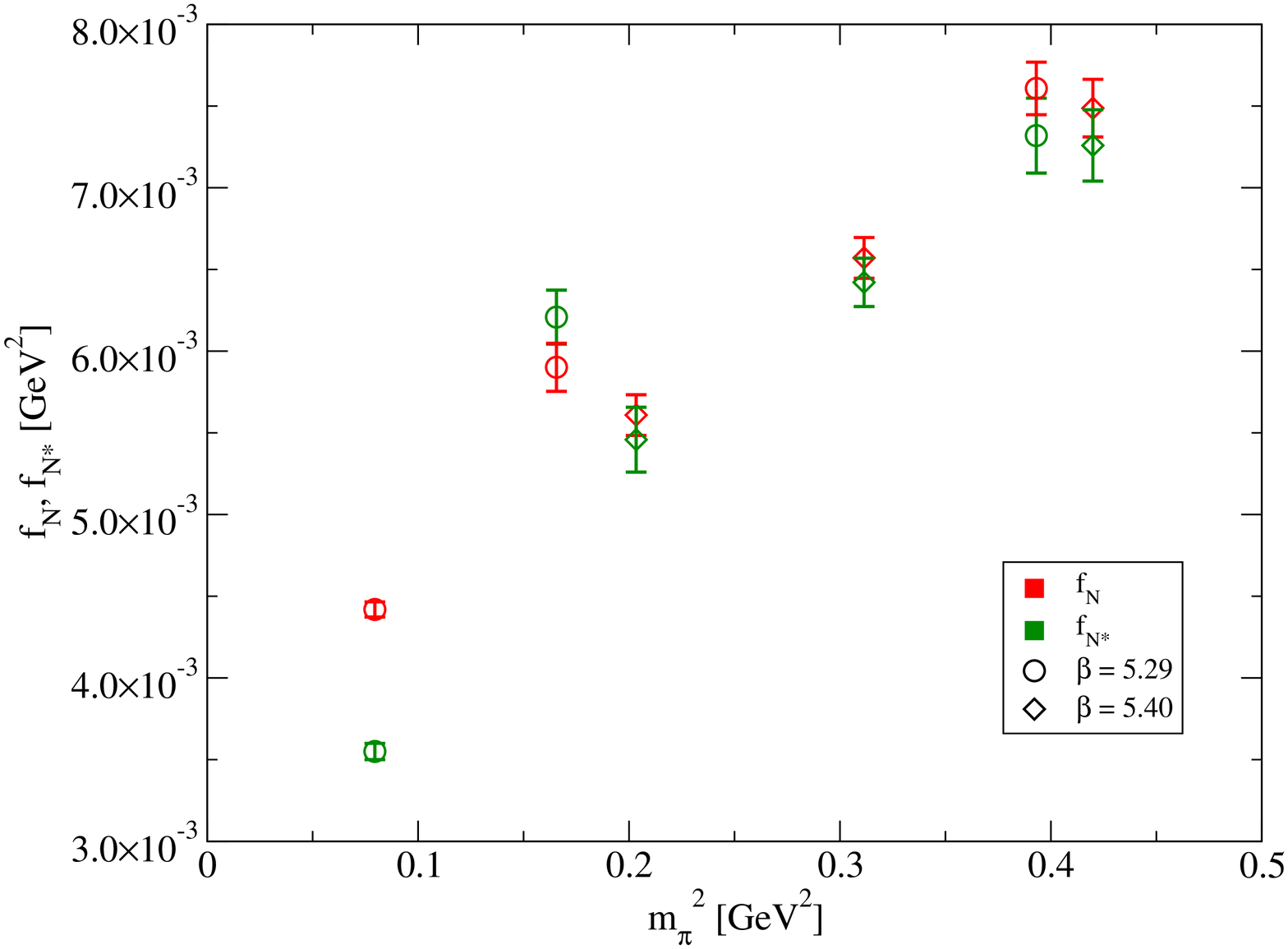}\\[-4.68cm]
\hspace*{5.55cm}
   \includegraphics[width=0.415\textwidth,angle=-90]{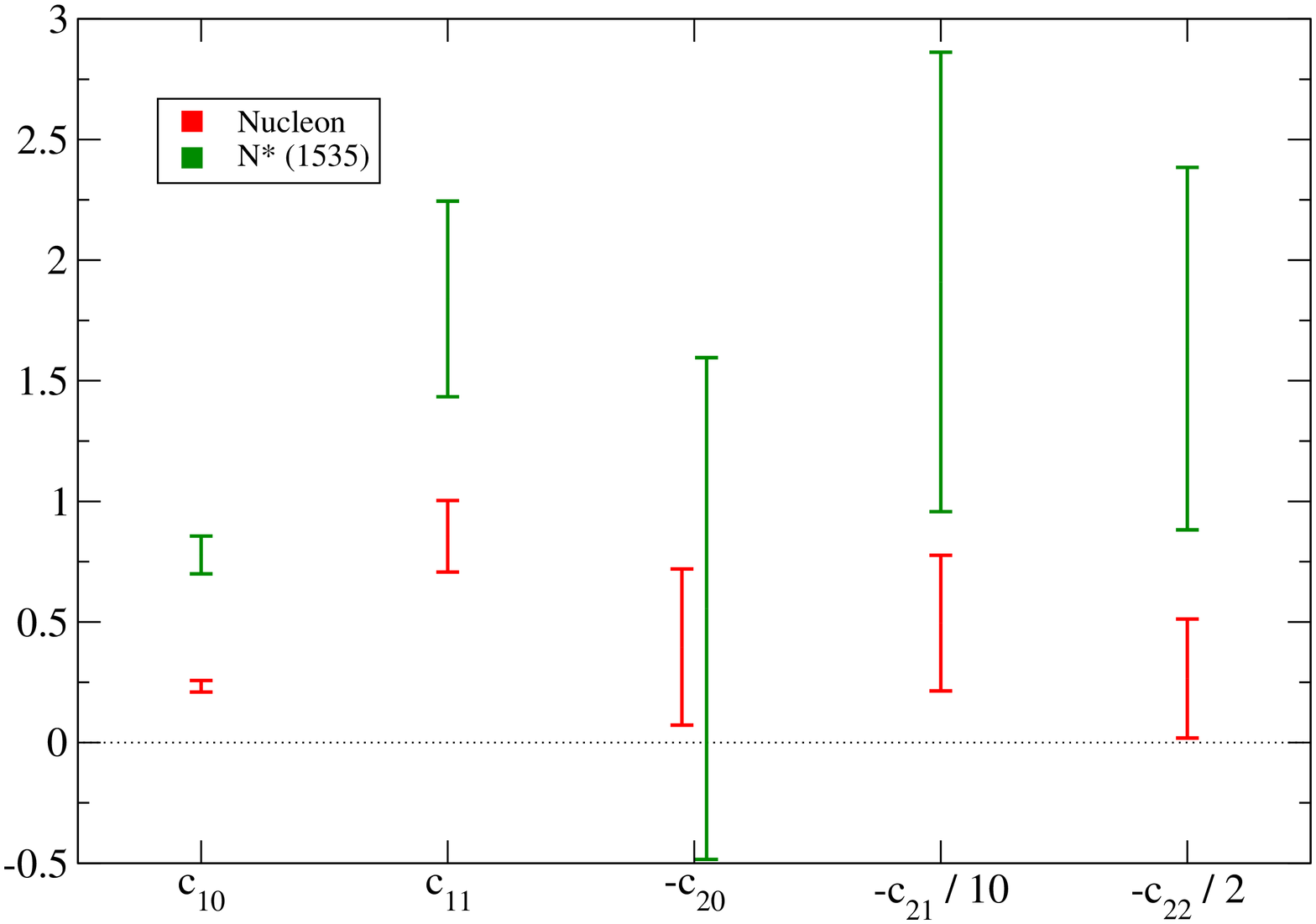}
\caption{Wave functions at the origin (left panel) and shape parameters of the nucleon 
        and $N^*(1535)$ distribution amplitudes (right panel).
         Only statistical errors are shown.}
\end{figure}
The left panel shows the comparison of nucleon and  $N^*(1535)$ wave functions at the 
origin, $f_N$ and $f_{N^*}$, as a function of $m_\pi^2$. 
The left-most points at $m_\pi^2 \simeq 0.08$~GeV$^2$ are new. They are much 
closer to the physical point and show, for the first time, that for small pion masses
$f_{N^*}$ becomes smaller than  $f_N$. This phenomenon still has to be understood.      

The results for shape parameters of the nucleon distribution amplitudes (\ref{twist3})
are shown on the right panel. One sees that coefficients of first-order polynomials 
$c_{10}$ and $c_{11}$ can be quantified, whereas  contributions of second order 
polynomials are less constrained. The main reason for this are the ${\mathcal O}(a)$
discretization errors in the chain rule for derivatives 
$D(A\cdot B) = (DA)\cdot B + A\cdot (DB) +{\mathcal O(a)}$ 
which spoil energy conservation: after the (nonperturbative) 
renormalization we obtain for the sum of the quark momentum 
fractions $x_1+x_2+x_3\simeq 0.94$ instead of unity. This is one of the issues that      
have to be addressed in future studies.

The main result so far is that the accuracy of modern lattice calculations is sufficient 
to detect differences in quark momentum fraction distributions in the nucleon and 
its parity partner state. Our calculations support a qualitative picture suggested 
by QCD sum rules\cite{Chernyak:1983ej} that the valence quark with the spin parallel 
to that of the nucleon carries most of its momentum, and for $N^*(1535)$ 
this effect appears to be even stronger. As an illustration we show in Fig.~2 the 
DAs of the nucleon and $N^*(1535)$ in barycentric coordinates.%
\footnote{In difference to the similar plot in Ref.~[\refcite{Braun:2009jy}] we only include 
the terms in $c_{10}$ and $c_{11}$ and discard contributions of second order polynomials 
which have large errors. }  
\begin{figure}[t]
\begin{center}
  \includegraphics[width=0.85\textwidth,angle=0]{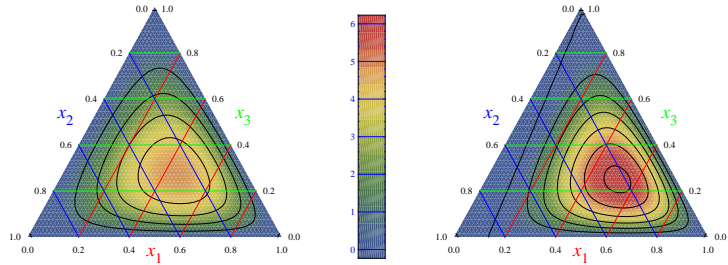}
\end{center}
\caption{Barycentric plots for the nucleon (left) and $N^*(1535)$ (right) 
distribution amplitudes.}
\end{figure}   
Note that the $N^*(1535)$ DA appears to be more narrow and shifted towards the lower-right
corner. 

All these results are preliminary and the study will be continued using larger
lattices and smaller pion masses. The comparison of the nucleon and $N^*(1535)$ will 
remain our primary goal for some time, but the calculations will also be extended to 
the full  $J^P=1/2^+$ and $J^P=1/2^-$ baryon octets and to the decuplets.
The main problem that has to be addressed in future is the 
identification of resonance contributions
for small quark (pion) masses such that strong decays, e.g. $N^*(1535)\to N\pi,N\eta$,
are allowed. The separation of $N^*(1535)$ and $N^*(1650)$ may prove to be difficult.
In particular the large decay width of $N^*(1535)$ in the $N\eta$ channel 
has to be understood and it can only be addressed in lattice calculations with three 
flavors of dynamic fermions. Discretization errors ${\mathcal O(a)}$ in the lattice 
definition of the relevant operators become a serious issue for second moments and
have to be reduced in order that the results for $c_{2k}$ coefficients become 
fully quantitative.  
  
\section{Transition form factors from Light-Cone Sum Rules}

The matrix element of the electromagnetic current $j^{\rm em}_\nu$ between spin-1/2
states of opposite parity can be parametrized in terms of two independent 
form factors, which can be chosen as
\begin{eqnarray} 
\label{Nstar_FF}
&&\langle N^*(P') | j_{\nu}^{\rm em} |N(P)\rangle \,=\, 
 \bar{u}_{N^*}(P')\gamma_5 \Gamma_\nu u_N(P)\,,
\nonumber\\
&& \Gamma_\nu \,=\,\frac{{G_1(q^2)}}{m_N^2}(\!\not\!q q_\nu - q^2 \gamma_\nu)
 -i \frac{{G_2(q^2)}}{m_N} \sigma_{\nu\rho} q^{\rho} \,,
\end{eqnarray}
where $q=P'-P$ is the momentum transfer.
The LCSRs are derived from the correlation function 
\[\int\! dx\, e^{-iqx}\langle N^*(P)| T \{ \eta (0) j_\mu^{\rm em} (x) \} 
| 0 \rangle \,,\]
where $\eta$ is a suitable operator with nucleon quantum numbers, e.g. 
the Ioffe current~\cite{Ioffe:1981kw}.
\begin{figure}[t]
\centerline{\epsfxsize4cm\epsffile{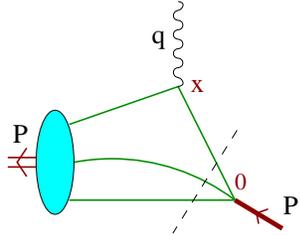}}
\caption{\label{figsum}\small
Schematic structure of the light-cone sum rule for baryon form factors.
}
\end{figure}
Making  use of the duality of QCD quark-gluon and hadronic degrees of freedom  
through dispersion relations one can write a representation for the 
form factors appearing in (\ref{Nstar_FF}) in terms of the DAs of $N^*$.
Schematically, the sum rules take the form
\[G_{1,2}(Q^2) = \sum_{k}\int[dx]\, C_{1,2;k}(x_i,Q^2,s_0,M^2,\mu,\alpha_s(\mu))\,\Phi_k(x_i,\mu) 
\] 
where the sum goes over contributions of nucleon DAs of increasing twist and 
$C_{k}$ are the coefficient functions that can be calculated in QCD perturbation 
theory and modified using dispersion relations to include two nonperturbative 
parameters: $s_0$, the interval of duality, and $M^2$, the Borel parameter which
specifies the distance (in imaginary time) on which matching between 
hadronic and quark representations for the correlation function is being done.
The dependence on $M^2$ is unphysical in the same sense as the factorization 
scale $\mu$ dependence in truncated perturbative expansions, but it is usually weak.
The pQCD limit\cite{Chernyak:1977as,Efremov:1979qk,Lepage:1979za,Lepage:1980fj}
corresponds to the leading part of the contribution of leading-twist DAs 
at large momentum transfer, and the 
main difference is that higher-twist contributions are {\it not} suppressed by powers
of $\Lambda^2_{\rm QCD}/Q^2$, but rather of $\Lambda^2_{\rm QCD}/s_0$ with 
$s_0\sim (1.5~$GeV$)^2$. The reason for this is that LCSRs take into account ``soft''
contributions to the form factors coming from large transverse separations. The
attractive feature of this approach is that such terms are calculated in terms 
of the DAs, thus avoiding the need to know (model) the full nonperturbative $k_\perp$
dependence of wave functions. 
       
In leading order, the  sum rules for $Q^2G_1(Q^2)/(m_N m_{N^*})$ and $-2G_2(Q^2)$
have the same functional form as the similar sum rules~\cite{Braun:2001tj,Braun:2006hz} 
for the Dirac and Pauli electromagnetic form factors of the proton,
with the replacement $m_N\to m_{N^*}$ in the light-cone expansion part, 
and different DAs. 

The experimental results for the electroproduction of $N^*(1535)$
are usually presented for helicity amplitudes  
$A_{1/2}(Q^2)$ and $S_{1/2}(Q^2)$ which 
can be expressed in terms of the form factors \cite{Aznauryan:2008us}:
\begin{eqnarray}
A_{1/2} &=& e\, B\, 
\Big[ Q^2 G_1(Q^2) + m_N(m_{N^*}-m_N)G_2(Q^2) \Big],
\nonumber \\
{S}_{1/2}&=&  \frac{e}{\sqrt{2}}B\,C\, 
\Big[(m_{N}-m_{N^*})G_1(Q^2)+m_{N}G_2(Q^2)\Big].
\nonumber
\end{eqnarray}
Here $e$ is the elementary charge and $B$, $C$ are kinematic factors defined as
\begin{eqnarray}
   B &=& \sqrt{\frac{Q^2+(m_{N*}+m_N)^2}{2 m_N^5(m_{N^*}^2-m_N^2)}} \,,
\qquad
   C  = \sqrt{1+\frac{(Q^2-m_{N^*}^2+m_N^2)^2}{4 Q^2 m_{N^*}^2}} \,.
\nonumber
\end{eqnarray}
\begin{figure}[t]
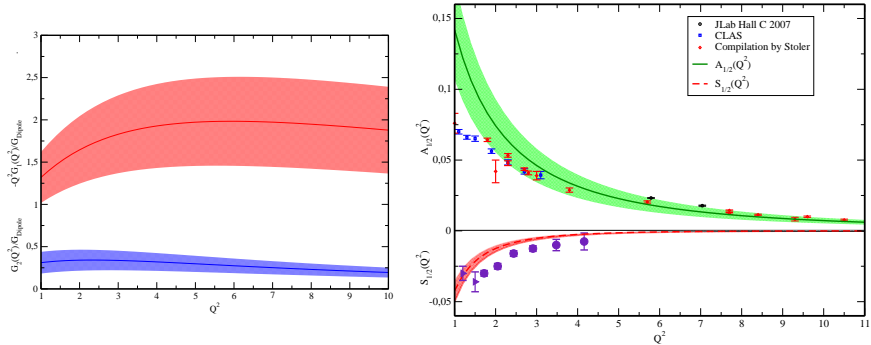

\vspace*{0.2cm}
  \includegraphics[width=0.450\textwidth,angle=0]{G12Errors.eps}\\[-4.2cm]
\hspace*{5.35cm}
   \includegraphics[width=0.530\textwidth,angle=0]{A12S12plusData.eps}
\caption{Form factors (left panel) and helicity amplitudes (right panel)
         for the electroexcitation of the $N^*(1535)$ resonance calculated 
         using the DAs in Ref.~[\refcite{Braun:2009jy}]. The experimental data 
         are from 
         Refs.~[\refcite{Aznauryan:2009mx,Dalton:2008ff,Denizli:2007tq,Stoler:1993yk}].}
\end{figure}

The results of the LCSR calculation of the form factors (normalized to the dipole) 
and helicity amplitudes using lattice-constrained $N^*$ DAs 
from Ref.~[\refcite{Braun:2009jy}] is presented in Fig.~4.  
The shaded areas show the estimated uncertainties. The $Q^2$-dependence of the form factors
$Q^2G_1(Q^2)$ and $G_2(Q^2)$ is predicted to be similar to the Dirac and Pauli nucleon
electromagnetic form factors, respectively. Only the normalization is different: 
The observed negative amplitude $S_{1/2}$ implies a small value of $G_2$.

These results have to be viewed as exploratory and further work is 
needed to make them quantitative. First of all, the LCSRs for 
baryon form factors have to be extended to the next-to-leading (NLO) accuracy,
i.e. including the $\mathcal{O}(\alpha_s)$ corrections. 
This calculation is rather cumbersome and not straightforward because of 
contributions of evanescent operators.
As the first step in this direction,
the NLO corrections to contributions of leading-twist DAs in the LCSRs for 
the electromagnetic nucleon form factors $F_1$ and $F_2$ have been calculated in 
Ref.~[\refcite{PassekKumericki:2008sj}], see Fig.~5. One sees that such corrections 
are significant, especially in the $G_E/G_M$ ratio. The extension of this calculation 
to contributions of sub-leading twist-4 DAs and the detailed study of 
baryon mass corrections to the sum rules is planned for the coming years.      
\begin{figure}[t]
\begin{center} 
\includegraphics[width=0.42\textwidth,angle=0]{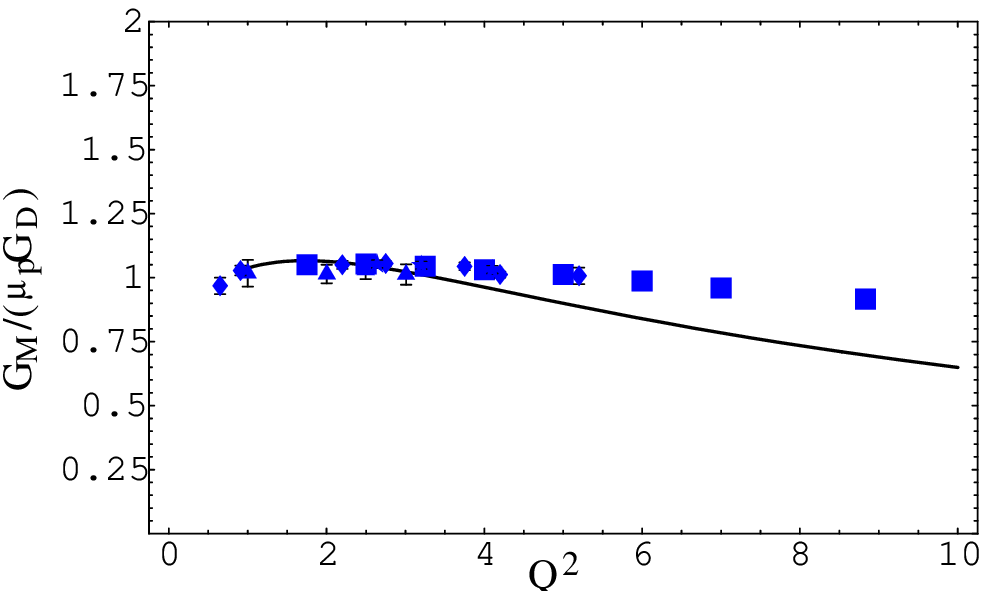}
~~~
\includegraphics[width=0.42\textwidth,angle=0]{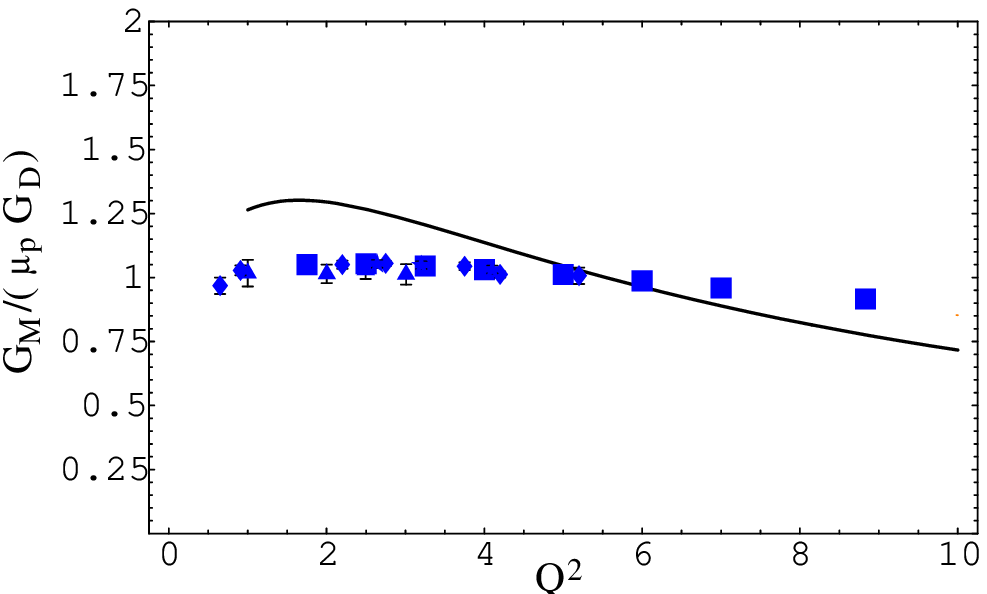}
\end{center}
\begin{center} 
\includegraphics[width=0.42\textwidth,angle=0]{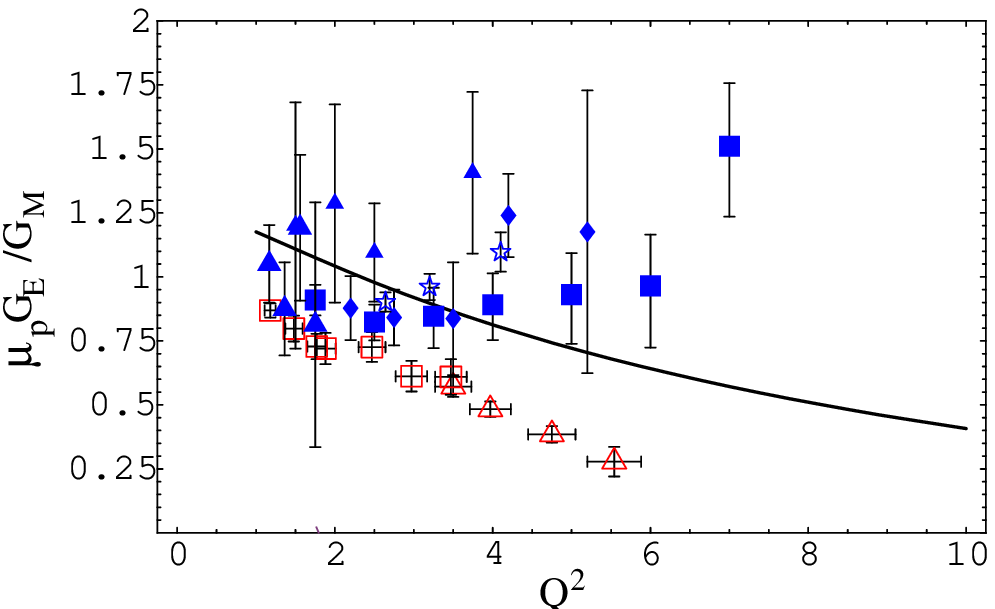}
~~~
\includegraphics[width=0.42\textwidth,angle=0]{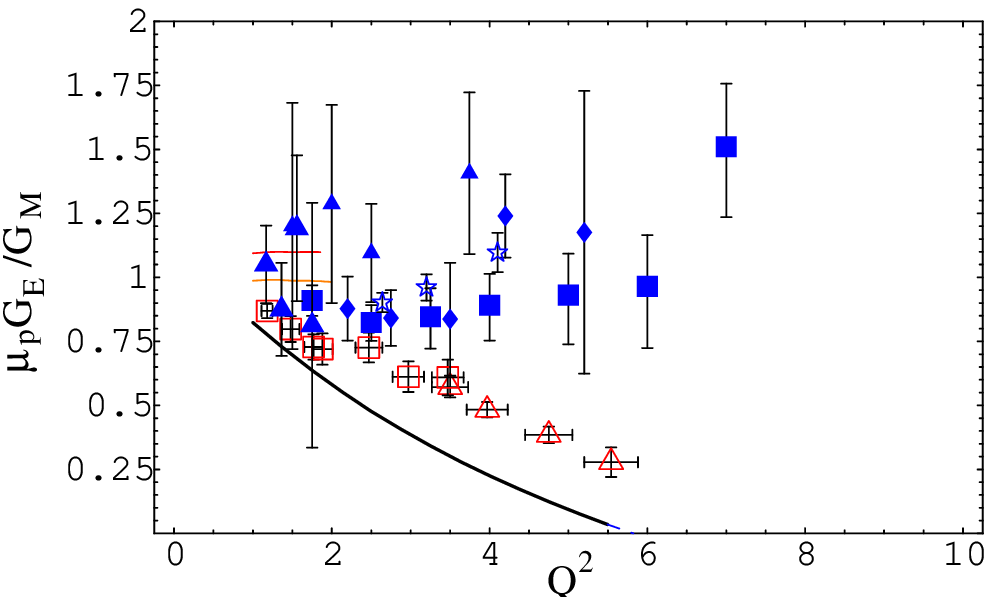}
\end{center}
\caption{\small LCSR results for the electromagnetic proton form factors 
for a realistic model of nucleon distribution amplitudes.
Left panel: Leading order (LO); right panel: 
next-to-leading order (NLO) for twist-three contributions.
Figure adapted from Ref.~[\refcite{PassekKumericki:2008sj}].}
\end{figure}

Second, one has to remember that the LCSRs suffer from irreducible uncertainties 
due to the duality assumption for contributions of higher resonances and the 
continuum which for meson form factors is estimated to be of order 10\%.
(see e.g. Ref.~[\refcite{Braun:1999uj}]). The experience
with applying this method to baryons is much less than for mesons, so that 
this ``systematic error'' cannot be estimated reliably. It is therefore imperative 
to apply the same technique to a maximally broad class of reactions.
At present applications include 
nucleon electromagnetic and axial form factors\cite{Braun:2001tj,Braun:2006hz,Wang:2006uva}, 
$N\gamma\Delta$ transitions\cite{Braun:2005be}, pseudoscalar- and vector-meson couplings 
to octet and decuplet baryons\cite{Aliev},
weak decays of the type $\Lambda_b\to p\ell\nu_\ell$\cite{Huang:2004vf}
and $\Lambda_{b}\to \Lambda \ell^+\ell^- $\cite{Aliev:2010uy}, 
and pion electroproduction at threshold\cite{Braun:2006td,Braun:2007pz}.
I expect that this list will continue to grow, giving us confidence in the whole
program.

\section*{Acknowledgments}
The author thanks 
M.~G{\"o}ckeler,      
R.~Horsley,          
T.~Kaltenbrunner,   
A.~Lenz,         
Y.~Nakamura,     
D.~Pleiter,         
P.~E.~L.~Rakow,         
J.~Rohrwild,       
A.~Sch{\"a}fer, 
R.~Schiel,      
G.~Schierholz,      
H.~St{\"u}ben,        
N.~Warkentin,     
J.~M.~Zanotti for the collaboration on this project.
Special thanks are due to Rainer~Schiel for the possibility to
use the results in Figs.~1,2 prior their presentation at 
LATTICE 2010\cite{schiel}.
This work is supported by the DFG through SFB/TR55 (Hadron Physics
from  Lattice QCD) and grants 9209070, 9209475.

\end{document}